\documentclass[aps,prl,twocolumn,superscriptaddress,floatfix]{revtex4-1}

\usepackage{palatino}
\usepackage{amsmath}
\usepackage{amssymb}
\usepackage{graphicx}
\usepackage{dcolumn}
\usepackage{boxedminipage}
\usepackage{verbatim}
\usepackage[colorlinks=true,citecolor=blue,linkcolor=blue]{hyperref}

\graphicspath{{../figures/}}

\newcommand{\expect}[1]{\left \langle #1 \right \rangle}                % <.> for denoting expectations over realizations of an experiment or thermal averages

\begin{document}

%%%%%%%%%%%%%%%%%%%%%%%%%%%%%%%%%%%%%%%%%%%%%%%%%%%%%%%%%%%%%%%%%%%%%%%%%%%%%%%%%%%%%%%%%%%%%%%%%%%%%%
% TITLE AND AUTHORS
%%%%%%%%%%%%%%%%%%%%%%%%%%%%%%%%%%%%%%%%%%%%%%%%%%%%%%%%%%%%%%%%%%%%%%%%%%%%%%%%%%%%%%%%%%%%%%%%%%%%%%

%\title{When is extension a good reaction coordinate in single-molecule pulling experiments?}
\title{Splitting probabilities as a test of reaction coordinate choice in single-molecule experiments}

 \author{John D. Chodera}
 \email{jchodera@berkeley.edu}
 \affiliation{California Institute of Quantitative Biosciences (QB3), University of California, Berkeley, CA 94720}
 
\author{Vijay S. Pande}
 \thanks{Corresponding author}
  \email{pande@stanford.edu}
 \affiliation{Department of Chemistry, Stanford University, Stanford, CA 94305}

\date{\today}

%%%%%%%%%%%%%%%%%%%%%%%%%%%%%%%%%%%%%%%%%%%%%%%%%%%%%%%%%%%%%%%%%%%%%%%%%%%%%%%%%%%%%%%%%%%%%%%%%%%%%%
% ABSTRACT/pacs
%%%%%%%%%%%%%%%%%%%%%%%%%%%%%%%%%%%%%%%%%%%%%%%%%%%%%%%%%%%%%%%%%%%%%%%%%%%%%%%%%%%%%%%%%%%%%%%%%%%%%%

\begin{abstract}
To explain the observed dynamics in equilibrium single-molecule measurements of biomolecules, the experimental observable is often chosen as a putative reaction coordinate along which kinetic behavior is presumed to be governed by diffusive dynamics. 
Here, we invoke the splitting probability as a test of the suitability of such a proposed reaction coordinate.  Comparison of the observed splitting probability with that computed from the kinetic model provides a simple test to reject poor reaction coordinates.  
We demonstrate this test for a force spectroscopy measurement of a DNA hairpin.
\end{abstract}

\pacs{PACS codes go here}

\maketitle

%%%%%%%%%%%%%%%%%%%%%%%%%%%%%%%%%%%%%%%%%%%%%%%%%%%%%%%%%%%%%%%%%%%%%%%%%%%%%%%%%%%%%%%%%%%%%%%%%%%%%%
% INTRODUCTION
%%%%%%%%%%%%%%%%%%%%%%%%%%%%%%%%%%%%%%%%%%%%%%%%%%%%%%%%%%%%%%%%%%%%%%%%%%%%%%%%%%%%%%%%%%%%%%%%%%%%%%

% Advances in experiment have allowed single-molecule measurements under equilibrium and nonequilibrium conditions.

A variety of new experimental techniques have made it possible to monitor the conformational fluctuations of single biological macromolecules under both equilibrium and nonequilibrium conditions.
These experiments aim to probe the statistical dynamics and conformational substates relevant to folding and function.
In a typical experiment, such as observation of the resonant energy transfer efficiency between two fluorophores incorporated into an RNA molecule~\cite{scherer:jmb:2008:equilibrium}, fluctuations of a spectroscopic observable in the absence of an external field are monitored.
Other experiments allow the effect of an external biasing potential on the dynamics to be observed, as in an optical trap~\cite{woodside:prl:2005:optical-force-clamp,collin:biophys-j:2006:tar-rna-hopping,woodside:science:2006:dna-hairpin-optical-trap}.

%Since the discovery of fundamental nonequilibrium work relations~\cite{jarzynski:prl:1997:nonequilibrium-work-relation,crooks:pre:1999:fluctuation-theorem}, a number of experiments that start in equilibrium but carry the system transiently out of equilibrium through the application of a time-dependent external force have been used to extract potentials of mean force along an extension coordinate~\cite{liphardt:science:2002:test-of-jarzynski-equality,bustamante:nature:2005:verification-of-crooks}.
%While these experiments have proven to be a powerful way to measure potentials of mean force, we shall hereafter restrict our discussion to equilibrium single-molecule experiments for simplicity.

% There is a desire to describe the observed dynamics along observed coordinate.

To describe the observed dynamics of the system, it is tempting to identify the observable with a reaction coordinate and construct a model in which the dynamics evolves by a diffusion process in an effective potential, such as by overdamped Langevin (also called ``Brownian'') dynamics~\cite{gardiner:handbook-of-stochastic-methods},
\color{black}
\begin{eqnarray}
\dot{x}(t) &=& - \beta \frac{\partial}{\partial x}F(x) + \sqrt{2 D(x)} R(t) \label{equation:brownian-equation-of-motion} .
\end{eqnarray}
Here, $x(t)$ is the time-dependent motion along the resolved coordinate, $D(x)$ is the diffusion constant (often assumed to be a constant independent of $x$), 
\color{black}
$\beta \equiv (k_B T)^{-1}$ is the inverse temperature, $F(x) \equiv - k_B T\ln \pi(x)$ is the potential of mean force (PMF) defined in terms of the observed equilibrium probability density $\pi(x)$, and $R(t)$ is a Gaussian process with zero mean satisfying $\expect{R(t) R(t')} = \delta(t - t')$.

Many physical systems such as biomolecules exhibit strong metastabilities in the conformational degrees of freedom, resulting in the presence of two or more discrete conformational states in which the system remains for a for long time before transitioning to another metastable state~\cite{schuette-huisinga:2003:metastable-states}.
While it is often easy to find an observable $x$ that is a suitable \emph{order parameter} that allows these metastable states to be discriminated to some degree, it is generally difficult to find a good \emph{reaction coordinate} so that dynamics along the resolved coordinate are well-described by Eq.~\ref{equation:brownian-equation-of-motion}.

%For a good choice of putative reaction coordinate $x$, there will exist a \emph{separation of timescales} in the motion along $x$ and motion orthogonal to it; the coordinate $x$ along which motion along the coordinate of interest can be effectively decoupled from the orthogonal degrees of freedom, which are subsumed into a memoryless stochastic bath whose effect is modeled by the stochastic process $R(t)$~\cite{hanggi-talkner-borkovec:1990:rev-mod-phys:fifty-years-after-kramers,zhou:quarterly-rev-biophys:2010:rate-theories-for-biologists}.
%On the other hand, a poor choice of reaction coordinate will result in a memory function that relaxes on the same order of magnitude as the correlation function.
%In this case, models that assume dynamics along the putative reaction coordinate can be described by a simple diffusion process such as Eq.~\ref{equation:brownian-equation-of-motion} will fail to adequately reproduce the statistics of the observed dynamics.

% How can we determine if a putative reaction coordinate is a good choice?

For data collected in a given single-molecule experiment, how can we determine whether the resolved coordinate provides a good reaction coordinate?
%Estimation of the memory function along the observed coordinate $x$ from observed timeseries data seems an obvious choice, as its relaxation behavior can be directly compared to the correlation function to evaluate the suitability of the observable as a reaction coordinate.
%While methods for estimating the memory function from observed trajectories date back to Berne and Harp~\cite{harp-berne:phys-rev-a:1970:estimating-memory-functions,harp-berne:adv-chem-phys::estimating-memory-functions} and Levesque and Verlet~\cite{levesque-verlet:phys-rev-a:1970:estimating-memory-functions}, these procedures often involve ill-conditioned numerical deconvolution or inversion of the Laplace transform, and newer approaches based on regularization~\cite{lange-grubmuller:jcp:2006:estimating-memory-functions} to make the inversion well-posed offer no assessment of statistical certainty of the estimated memory function.
Recent work on tests of reaction coordinate suitability in computer simulations has focused on the calculation of the \emph{committor} or \emph{splitting probabilities}, a concept dating back to Onsager~\cite{lange-grubmuller:jcp:2006:estimating-memory-functions}.
This quantity, now extensively used in simulation studies of protein folding~\cite{du-pande:jcp:1998:pfold}, represents the probability that a trajectory first encounters one absorbing boundary placed along the reaction coordinate before another, given an initial microscopic state of the system.
For suitable choices of reaction coordinate, the distribution of committor probabilities along an equilibrium ensemble of configurations restricted to a given value of the reaction coordinate will be closely grouped about a characteristic value~\cite{du-pande:jcp:1998:pfold,geissler:jpcb:1999:ion-dissociation,bolhuis:annu-rev-phys-chem:2002:transition-path-sampling,ma-dinner:jpcb:2005:automatic-reaction-coordinate-identification,peters:jcp:2006:histogram-test,peters:2010:cpl:committor}; indeed, ideal reaction coordinates organize committor isosurfaces in an ordered fashion along the reaction coordinate~\cite{rhee:jpcb:2005:splitting-probability}.
%for example, at the separatrix, from which trajectories have equal probability of reaching either absorbing boundary, the committor distribution should be closely centered about the value $0.5$.

Unfortunately, tests based on evaluating distributions of committor values along cuts of the putative reaction coordinate are impossible to apply in a physical experiment, since there is no way to prepare the system in precisely the same microscopic configuration to probe the statistics of committor probabilities.
Instead, we propose a simple alternative that is readily computable from observed equilibrium trajectories of the resolved coordinate: Comparison of the \emph{average} committor along each value of the reaction coordinate evaluated by Eq.~\ref{equation:brownian-equation-of-motion} with the \emph{empirical average} committor from the observed trajectory.

\noindent\emph{Theory.} 
Consider the placement of absorbing boundaries at $a$ and $b$ near the periphery of the observed range of the resolved coordinate $x$.
For a diffusion process in one dimension governed by Eq.~\ref{equation:brownian-equation-of-motion}, the probability of first encountering $a$ before $b$ starting from $x \in [a,b]$ can be shown to be~\cite{gardiner:handbook-of-stochastic-methods,rhee:jpcb:2005:splitting-probability,best-hummer:2010:pnas:coordinate-dependent-diffusion},
\color{black}
\begin{eqnarray}
p_A(x) &=& \frac{\int_{x}^{b} dx' \, D(x')^{-1} \, e^{\beta F(x')}}{\int_{a}^{b} dx' \, D(x')^{-1} \, e^{\beta F(x')}} \label{equation:splitting-probability-from-pmf} .
\end{eqnarray}
\color{black}
The PMF along the resolved coordinate $x$, $F(x)$, can be estimated from a single-molecule trajectory of sufficient length~\cite{woodside:science:2006:dna-hairpin-optical-trap,rief:2010:pnas:gcn4-pmf} or from multiple trajectories under different equilibrium~\cite{shirts-chodera:jcp:2008:mbar} or nonequilibrium~\cite{minh-adib:prl:2008:bidirectional-pulling,minh-chodera:jcp:2009:bidirectional-pulling} conditions.
\color{black}
The diffusion profile $D(x)$ can be estimated in a number of ways (such as the Bayesian scheme of Best and Hummer that allows simultaneous computation of both PMF and diffusion constant~\cite{best-hummer:2010:pnas:coordinate-dependent-diffusion}), though it is commonly assumed to be constant, in which case it cancels from both numerator and denominator.
\color{black}
%{\color{black}[JDC: Should we remove extraneous references and just cite Best-Hummer paper as a way to estimate $F(x)$ and $D(x)$ simultaneously?]}

An \emph{empirical} estimate of the splitting probability $\hat{p}_A(x)$ can also be computed directly from an observed equilibrium trajectory $x(t)$, 
\color{black}
$t \in [0,\mathcal{T}]$ (also recently noted in Ref.~\cite{thirumalai:2011:prl:pfold}),
\begin{eqnarray}
\hat{p}_A(y) &=& \frac{\int_{0}^{\mathcal{T}} dt \, \delta(y - x(t)) \, c_A(t)}{\int_{0}^{\mathcal{T}} dt \, \delta(y - x(t))} \label{equation:empirical-splitting-probability}
\end{eqnarray}
where we have defined the hitting function $c_A(t)$ in terms of $x(t)$ as,
\begin{eqnarray}
c_A(t) &=& \begin{cases}
1 & \mathrm{if} \:\: \tau_A(t) < \tau_B(t) \\
0 & \mathrm{otherwise}
\end{cases} ,
\end{eqnarray}
where auxiliary functions $\tau_A(t)$ and $\tau_B(t)$ are defined as,
\begin{eqnarray}
\tau_A(t) &=& \inf \{ t' > t : x(t) < a \} \nonumber \\
\tau_B(t) &=& \inf \{ t' > t : x(t) > b \} .
\end{eqnarray}
\color{black}
The hitting function $c_A(t)$ simply keeps track of whether $x(t)$ will hit boundary $a$ before $b$ immediately following time $t$, and assumes the value of unity if so, and zero otherwise.
In practice, the delta function $\delta(y - x(t))$ is replaced by some kernel function of finite width, such as a histogram bin.
An estimate of $\hat{p}_A(x)$ from multiple equilibrium trajectories can be produced by averaging the trajectories weighted by their lengths.

\begin{figure}[tb]
\resizebox{\columnwidth}{!}{\includegraphics{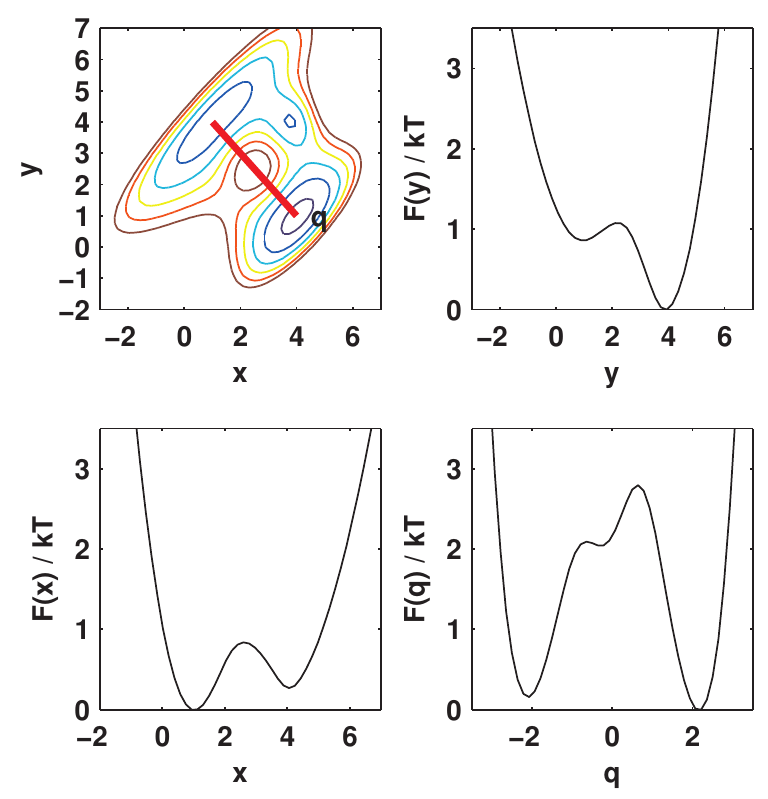}}
\caption{\label{figure:model-pmfs} {\bf Two-dimensional model system and potentials of mean force.}
\emph{Upper left:} Potential for the two-dimensional model system, with contours drawn every 5 $k_B T$.
$x$ and $y$ are poor reaction coordinates, while $q = (x - y) / \sqrt{2}$ (thick black line) is a good reaction coordinate.
\emph{Other panels:} Potentials of mean force in units of $k_B T$ for projections onto $x$, $y$, and $q$.} 
\end{figure}

Our proposed test is simple:
By comparing the splitting probability estimated from the PMF, $p_A(x)$, with the empirical estimate of the splitting probability from the trajectory, $\hat{p}_A(x)$, we can judge whether these quantities are obviously discrepant over the range $x \in [a, b]$, which would indicate that $x$ is a poor reaction coordinate.
Note that this test is \emph{necessary}, but not \emph{sufficient}, for $x$ to be a good reaction coordinate; agreement does not mean that the putative reaction coordinate is a true reaction coordinate.
Nevertheless, the test may be sufficiently exacting to reject poor choices of reaction coordinate that are not immediately obvious by eye yet fail this comparison.

\color{black}
When the observed coordinate is determined to be a poor reaction coordinate, the consequences of assuming it to be an adequate reaction coordinate depend on the precise nature of the information extracted from the single-molecule data.
The consequences could be as simple as underestimating the rate constant for a two-state process or as subtle as inferring an erroneous mechanism for more complex processes.
The most obvious consequence is mistaking the location of the transition state---the point where the splitting probability $p_A = 0.5$---to be displaced from the free energy barrier in the potential of mean force.
For systems like DNA hairpins and proteins, this can have consequences for the interpretation of how ``brittle'' or ``compliant'' the conformational states are perceived to be.
Notably, similar tests have been found to be useful in validating putative reaction coordinate choices in computer simulations, despite the ability to inspect the atomic coordinates directly~\cite{peters:jcp:2007:reaction-coordinate-test,bolhuis:jcp:2010:reaction-coordinate-test}.
\color{black}

%%%%%%%%%%%%%%%%%%%%%%%%%%%%%%%%%%%%%%%%%%%%%%%%%%%%%%%%%%%%%%%%%%%%%%%%%%%%%%%%%%%%%%%%%%%%%%%%%%%%%%
% MODEL SYSTEM
%%%%%%%%%%%%%%%%%%%%%%%%%%%%%%%%%%%%%%%%%%%%%%%%%%%%%%%%%%%%%%%%%%%%%%%%%%%%%%%%%%%%%%%%%%%%%%%%%%%%%%

\noindent\emph{Model system.}
As an illustrative example, we consider the two-dimensional model system previously studied by Rhee and Pande~\cite{rhee:jpcb:2005:splitting-probability},
\begin{eqnarray}
U(x,y) &=& [1 - 0.5 \tanh (y-x)] (x+y-5)^2 \\
&+& 0.2[((y-x)^2-9)^2+3(y-x)] \nonumber \\
&+& 15 e^{-(x-2.5)^2-(y-2.5)^2} - 20 e^{-(x-4)^2-(y-4)^2} \nonumber ,
\end{eqnarray}
pictured here in the upper-left panel of Fig.~\ref{figure:model-pmfs}.
Two stable states are present, located roughly at $(x,y)$-coordinates $(4,1)$ and $(1,4)$.
At $k_B T = 5$, the PMFs along both $x$ and $y$ clearly show two distinct wells separated by a barrier, and yet these coordinates are expected to be poor reaction coordinates individually; the coordinate $q = (x-y)/\sqrt{2}$ (Fig.~\ref{figure:model-pmfs}, upper-left panel, red line), however, which connects the two stable basins more directly, is known to be a good reaction coordinate at this temperature~\cite{rhee:jpcb:2005:splitting-probability}.

A Brownian dynamics trajectory of $10^6$ steps was generated using the discretization of Eq.~\ref{equation:brownian-equation-of-motion} by Ermak and Yeh~\cite{ermak-yeh:cpl:1974:brownian-dynamics,ermak:jcp:1975:brownian-dynamics}, with a diffusion constant of $D = 1$ and timestep $\Delta t = 0.1$.
This trajectory was projected onto either poor choices of reaction coordinate $x$ and $y$, or good reaction coordinate $q$ (Supplementary Fig.~1).
%For each resolved coordinate, the Bayesian method of Best and Hummer~\cite{best-hummer:2010:pnas:coordinate-dependent-diffusion} was used to estimate potential of mean force $F(x)$ and position-dependent diffusion constant $D(x)$ and associated confidence intervals on 60 equally-spaced bins.
For each projection, the potential of mean force was estimated from an empirical histogram, e.g.~$F(x) \approx - k_B T \ln p(x)$ for the projection onto $x$, using 100 equally-sized bins.
The PMF-derived splitting probability $p_A(x)$ was computed from $F(x)$ using Eq.~\ref{equation:splitting-probability-from-pmf}, and the empirical splitting probability $\hat{p}_A(x)$ according to Eq.~\ref{equation:empirical-splitting-probability}.
To judge whether disagreement between these estimates was statistically meaningful, the statistical uncertainty in the empirical $\hat{p}_A(x)$ was estimated using by time-correlation analysis (see Supplementary Information).
%{\color{black}[JDC: If we switch to Best and Hummer method, I'll just use their Bayesian scheme to estimate $F(x)$ and $D(x)$, and $p_A(x)$ and confidence intervals simultaneously.]}

%\begin{figure}[tb]
%\resizebox{\columnwidth}{!}{\includegraphics{rhee_pande_trajectory.pdf}}
%\caption{\label{figure:model-trajectories} {\bf Trajectory from model system projected onto poor and good choices of reaction coordinate.}
%A Brownian dynamics trajectory of $10^6$ steps is projected onto poor choices of reaction coordinate $x$ (top) and $y$ (middle), as well as $q = (y - x)/\sqrt{2}$.}
%\end{figure}

\begin{figure}[tb]
\resizebox{0.32\columnwidth}{!}{\includegraphics{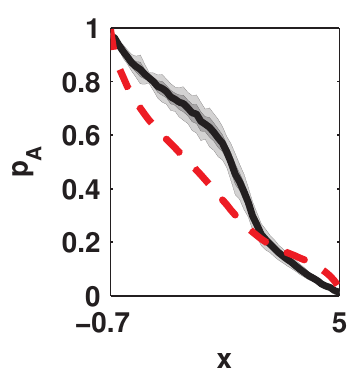}}
\resizebox{0.32\columnwidth}{!}{\includegraphics{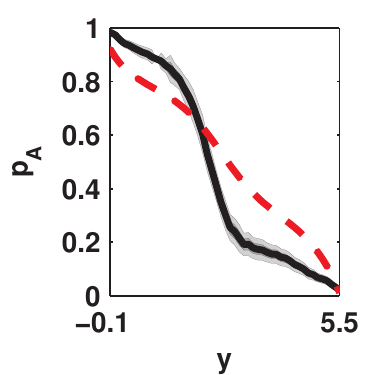}}
\resizebox{0.32\columnwidth}{!}{\includegraphics{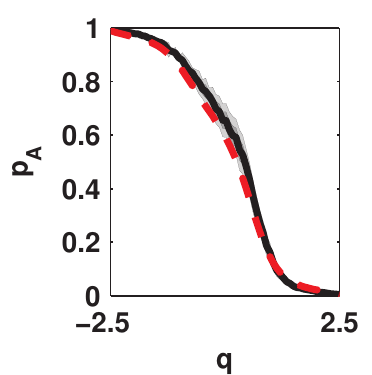}}
\caption{\label{figure:model-splitting-comparison} {\bf Splitting probability tests for two-dimensional model system.}
For each choice of projected coordinate shown in Fig.~\ref{figure:model-pmfs}, both the trajectory-derived empirical splitting probability $\hat{p}_A$ (solid black line) and the PMF-derived splitting probability $p_A$ (dashed black line) are shown.
%Projections onto $x$ (left) and $y$ (middle) are poor reaction coordinates, while projection onto $q$ (right) is a good reaction coordinate.
Dark shading represents a 68\% confidence interval about $\hat{p}_A$, and light shading a 95\% confidence interval.
}
\end{figure}

The results of this comparison {\color{black}assuming a uniform diffusion constant} are shown in Fig.~\ref{figure:model-splitting-comparison}.
The poor suitability of $x$ and $y$ as reaction coordinates is easily seen by the large discrepancy between the the splitting probability $p_A$ computed from the PMF (dashed line) and the empirical splitting probability $\hat{p}_A$ estimated from the trajectories (solid line).
However, the coordinate $q = (x - y)/\sqrt{2}$, previously identified by Rhee and Pande as being well-aligned with the true reaction coordinate at this temperature by sophisticated means not available to single-molecule experiments~\cite{rhee:jpcb:2005:splitting-probability}, agrees to within statistical error (shaded region).

%%%%%%%%%%%%%%%%%%%%%%%%%%%%%%%%%%%%%%%%%%%%%%%%%%%%%%%%%%%%%%%%%%%%%%%%%%%%%%%%%%%%%%%%%%%%%%%%%%%%%%
% APPLICATION TO EXPERIMENTAL DATA
%%%%%%%%%%%%%%%%%%%%%%%%%%%%%%%%%%%%%%%%%%%%%%%%%%%%%%%%%%%%%%%%%%%%%%%%%%%%%%%%%%%%%%%%%%%%%%%%%%%%%%

\begin{figure*}[tbp]
\noindent
%\resizebox{0.81\textwidth}{!}{\includegraphics{dna-hairpin.pdf}} \resizebox{0.18\textwidth}{!}{\includegraphics{best-hummer-extension.pdf}}
\resizebox{\textwidth}{!}{\includegraphics{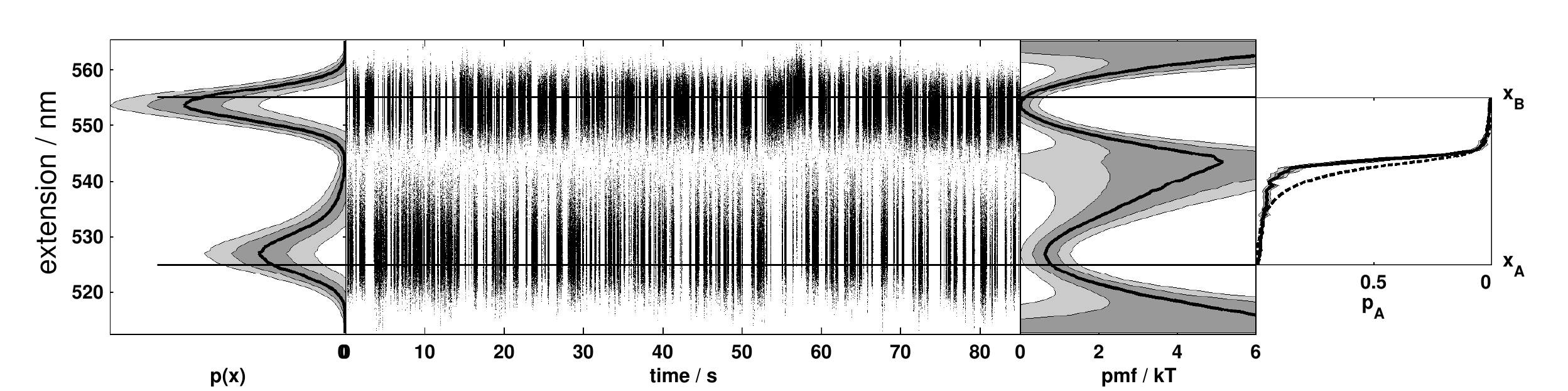}}
\caption{\label{figure:dna-hairpin} {\bf Splitting probability analysis for a DNA hairpin in a passive all-optical constant force double trap.}
From left to right: Histogram of observed values of the extension coordinate; 
complete observed trajectory of extension coordinate over experimental timecourse; 
potential of mean force along extension coordinate estimated from histogram; 
splitting probabilities estimated directly from trajectory (solid line) and computed from the potential of mean force (dashed line) using Eq.~\ref{equation:splitting-probability-from-pmf}.
Dark shaded regions around solid lines represent a 68\% symmetric confidence interval, and light shaded regions 95\% confidence interval.
Note that the bead-to-bead DNA hairpin extension coordinate (along the ordinate) is the same throughout all panels. 
} 
\end{figure*}
% TODO for figure:
% - Get rid of lines around shaded regions
% - Add dashed line across entire figure at x_A and x_B
% - Add ticks for x_A and x_B at far right, labeling them 'x_A' and 'x_B' instead of numbers
% - Change axis labels to have units read like "extension / nm", "time / s", and "f / kT"
% - Add p_B to top of splitting probability figure
% - Add zoomed-in view of some trajectory data at high time-resolution as inset.

\noindent\emph{DNA hairpin force spectroscopy.}
To demonstrate the utility of our proposed splitting probability test in a real laboratory measurement, we performed the same analysis on a single-molecule trajectory of a DNA hairpin in a double optical trap, previously reported by Woodside et al.~\cite{woodside:science:2006:dna-hairpin-optical-trap}.
The hairpin, referred to as 30R50/T4 due to the content of a 30 bp stem-forming sequence, is attached by means of dsDNA handles to two polystyrene beads held in a passive all-optical constant-force clamp~\cite{woodside:prl:2005:optical-force-clamp} at an external force that encourages hopping among closed and open conformations over the course of the experiment.
% F1/2 = 14.4$\pm$0.7 pN --- near here?
Bead displacements in the trap were recorded with a sampling frequency of 25 kHz~\cite{woodside:science:2006:dna-hairpin-optical-trap}, and the bead-to-bead extension trajectory was analyzed here.

Fig.~\ref{figure:dna-hairpin} shows the observed trajectory of the molecular extension coordinate and corresponding splitting probability analysis
\color{black}
for a uniform diffusion constant.
\color{black}
From this analysis, it is evident there is poor agreement between $p_A(x)$ estimated from the PMF and the empirical $\hat{p}_A(x)$ estimated from the trajectory in the region of extensions between 535 and 545 nm.
This suggests that, at this external force, dynamics would be poorly-described by Brownian dynamics along the total molecular extension coordinate using Eq.~\ref{equation:brownian-equation-of-motion} {\color{black} and a uniform diffusion constant}.

\begin{figure}[tbp]
\resizebox{0.49\columnwidth}{!}{\includegraphics{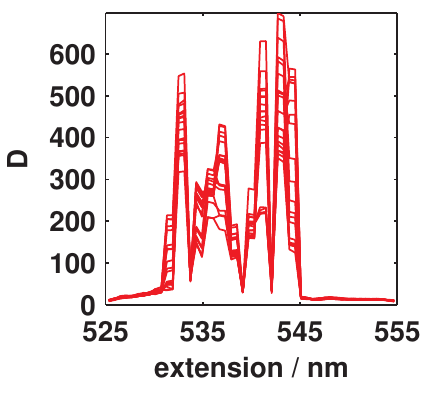}}
\resizebox{0.49\columnwidth}{!}{\includegraphics{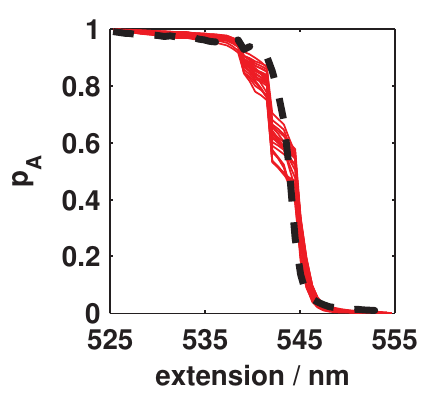}}
\caption{{\bf Position-dependent diffusion constant and splitting probability test incorporating position-dependent diffusion for DNA hairpin.}
\emph{Left:} Position-dependent diffusion constant, in nm/s$^2$.
\emph{Right:} Splitting probability test incorporating position-dependent diffusion constant, with empirical splitting probability $\hat{p}_A$ shown as a thick dashed line.
Because the Bayesian scheme of Best and Hummer~\cite{best-hummer:2010:pnas:coordinate-dependent-diffusion} was used to compute potentials of mean force and diffusion constants, the estimated diffusion constant $D(x)$ and PMF-derived splitting probabilities $p_A$ are shown as thin solid lines representing 20 samples from the Bayesian posterior.
}
\label{figure:coordinate-dependent-diffusion-hairpin}
\end{figure}

\color{black}
\emph{Non-uniform diffusion.}  Recently, it has been suggested that non-uniformity of the diffusion constant along the resolved coordinate may have important ramifications for single-molecule biophysical experiments~\cite{best-hummer:2010:pnas:coordinate-dependent-diffusion}.
Could strong position-dependence of the diffusion constant $D(x)$ may be responsible for the observed discrepancy in in Fig.~\ref{figure:dna-hairpin}?
To judge whether non-uniform diffusion significantly impacted our test of reaction coordinate suitability, we used the Bayesian inference scheme proposed by Best and Hummer~\cite{best-hummer:2010:pnas:coordinate-dependent-diffusion} to simultaneously compute position-dependent diffusion constant $D(x)$ and potential of mean force $F(x)$ for the systems considered here (Supplementary Figs.~2 and 3).
Notably, the diffusion constant varies markedly with the bead-to-bead extension (Fig.~\ref{figure:coordinate-dependent-diffusion-hairpin}, left), and the agreement of the PMF-derived $p_A$ and empirical $\hat{p}_A$ (Fig.~\ref{figure:coordinate-dependent-diffusion-hairpin}, right) improves substantially.
By contrast, repeating the reaction coordinate test for the 2D model system allowing for a position-dependent diffusion constant reveals only relatively minor variations in the estimated diffusion constant that result in no substantial change in which reaction coordinates are rejected by the test (Supplementary Fig.~2).
Taken together, these data suggest a significant role for position-dependent diffusion in the DNA hairpin system under force, in agreement with the theoretical findings of Best and Hummer~\cite{best-hummer:2010:pnas:coordinate-dependent-diffusion}.
\color{black}

%%%%%%%%%%%%%%%%%%%%%%%%%%%%%%%%%%%%%%%%%%%%%%%%%%%%%%%%%%%%%%%%%%%%%%%%%%%%%%%%%%%%%%%%%%%%%%%%%%%%%%
% DISCUSSION
%%%%%%%%%%%%%%%%%%%%%%%%%%%%%%%%%%%%%%%%%%%%%%%%%%%%%%%%%%%%%%%%%%%%%%%%%%%%%%%%%%%%%%%%%%%%%%%%%%%%%%

%\begin{itemize}
%  \item Test only necessary, not sufficient.
%  \item Even if good reaction coordinate at measured force, don't know if will be good at low force; maybe Iaon Ianescu work will help there?
%  \item Quality of reaction coordinate is really a matter of degree
%\end{itemize}

\noindent\emph{Discussion.}
We note that the reaction coordinate test presented here only allows us to test a condition that is \emph{necessary}, but not \emph{sufficient}, for Brownian dynamics to appropriately describe the observed dynamics on a one-dimensional landscape determined by the PMF.
\color{black}
This does not rule out the possibility of pathological cases where poor reaction coordinates go unnoticed because the average splitting probability at a particular value of the resolved coordinate matches the PMF-derived model, but the splitting probability distribution is not tightly peaked about its average value.
Additionally, if multiple reactive channels exist that are otherwise indistinguishable by this test, differences between the channels will not be resolvable.

Despite this, our test was able to discern good from poor choices of reaction coordinate in a model system, and reject the extension coordinate as a good choice of coordinate for a DNA hairpin unless a strongly position-dependent diffusion constant is permitted.
Even then, there are statistically significant discrepancies between the observed splitting probability and the PMF-derived splitting probability that indicate this reaction coordinate choice is not ideal. 
We note that the presence of $\sim$ 1 kb dsDNA handles tethering the DNA hairpin to the laser-trapped polystyrene beads is one potential source of the incomplete alignment of the extension coordinate with the reaction coordinate for hairpin unzipping.
Shorter dsDNA handles have recently been suggested as a way to improve the signal-to-noise ratio~\cite{ritort:biophys-j:2011:short-handles}, and may also improve the reaction coordinate quality.
For proteins, techniques that allow the attachment of tethers at specific attachment points can be exploited to probe for improved reaction coordinate should the experimenter find that the current pulling coordinate under study is unsuitably poor~\cite{marqusee-bustamante:eur-biophys-j:2008:protein-optical-tweezers}.
\color{black}
Finally, we note that though this test is able to test the suitability of the extension coordinate for a polymer under force, we cannot determine from the present analysis whether a good reaction coordinate in the presence of external force would also be a good reaction coordinate in the \emph{absence} of force, or even under different biasing forces; this concern is still the subject of active study~\cite{nummela-andricioaei:2007:biophys-j:low-force-kinetics,best-paci-hummer-dudko:2008:jpcb:pulling-reaction-coordinate}.

%{\color{black}[JDC: Additional papers we should work in citations for:
%%Estimation of PMF from single-molecule data (Woodside~\cite{woodside:science:2006:dna-hairpin-optical-trap}, Shirts~\cite{shirts-chodera:jcp:2008:mbar}, Rief~\cite{rief:2010:pnas:gcn4-pmf});
%%Extracting low-force kinetics from high-force measurements~\cite{nummela-andricioaei:2007:biophys-j:low-force-kinetics};
%%Universality of splitting probability as reaction coordinate~\cite{rhee:jpcb:2005:splitting-probability};
%%Pulling coordinate as a reaction coordinate~\cite{best-paci-hummer-dudko:2008:jpcb:pulling-reaction-coordinate};
%%Coordinate-dependent diffusion~\cite{best-hummer:2010:pnas:coordinate-dependent-diffusion};
%%Baron Peters~\cite{peters-trout:2006:jcp:likelihood-method,peters:2006:jcp:histogram-test,peters:2010:cpl:committor};
%Antoniou~\cite{antoniou-schwartz:2009:jcp:stochastic-separatrix};
%Szabo~\cite{berezhkovskii-szabo:2005:jcp:reaction-coordinates};
%%Fifty years after Kramers~\cite{hanggi-talkner-borkovec:1990:rev-mod-phys:fifty-years-after-kramers};
%Krivov and Karplus, progress variables that preserve barriers~\cite{krivov-karplus:2006:jpcb:progress-variables};
%Thirumalai, transition state from single-molecule experiments~\cite{thirumalai:2011:prl:pfold} .
%]}

%%%%%%%%%%%%%%%%%%%%%%%%%%%%%%%%%%%%%%%%%%%%%%%%%%%%%%%%%%%%%%%%%%%%%%%%%%%%%%%%%%%%%%%%%%%%%%%%%%%%%%
% ACKNOWLEDGMENTS
%%%%%%%%%%%%%%%%%%%%%%%%%%%%%%%%%%%%%%%%%%%%%%%%%%%%%%%%%%%%%%%%%%%%%%%%%%%%%%%%%%%%%%%%%%%%%%%%%%%%%%
\begin{acknowledgments}
\noindent\emph{Acknowledgments.}
The authors thank Michael Woodside (University of Alberta and National Institute for Nanotechnology, NRC), Phillip Elms and David Chandler (U Berkeley), Gerhard Hummer and Attila Szabo (NIH), Steven Block and Imran Haque (Stanford University), Felix Ritort (University of Barcelona), {\color{black} and the anonymous referees} for their helpful feedback on this work.
The authors are grateful to Michael Woodside and Steven M.~Block (Stanford University) for kindly providing original single-molecule data.
JDC acknowledges support through an NSF grant for Cyberinfrastructure (NSF CHE-0535616) and a California Institute of Quantitative Biosciences (QB3) Distinguished Postdoctoral Fellowship.
VSP acknowledges support from NIH R01-GM062868, NSF-DMS-0900700, NSF-MCB-0954714, and NSF EF-0623664.
Matlab code implementing the analysis procedure described here can be obtained from \url{https://simtk.org/home/splitting}.
\end{acknowledgments}

%%%%%%%%%%%%%%%%%%%%%%%%%%%%%%%%%%%%%%%%%%%%%%%%%%%%%%%%%%%%%%%%%%%%%%%%%%%%%%%%%%%%%%%%%%%%%%%%%%%%%%
% BIBLIOGRAPHY
%%%%%%%%%%%%%%%%%%%%%%%%%%%%%%%%%%%%%%%%%%%%%%%%%%%%%%%%%%%%%%%%%%%%%%%%%%%%%%%%%%%%%%%%%%%%%%%%%%%%%%
\bibliography{single-molecule-pfold}

\end{document}